\documentstyle[12pt]{article}

\begin{document}

\thispagestyle{empty}

\begin{flushright}
{\parbox{3.5cm}{
UAB-FT-358

hep-ph/9502317

January, 1995
}}
\end{flushright}

\vspace{1cm}
\hyphenation{ne-ver-the-less}
\hyphenation{cano-nical ca-nonical canoni-cal}
\begin{center}
\begin{large}
\begin{bf}
THE QUANTUM CORRELATION $R_b$-$R_c$ IN THE MSSM: MORE HINTS OF SUPERSYMMETRY?\\
\end{bf}
\end{large}
\vspace{1.25 cm}
David GARCIA
\footnote{Internet address:GARCIA@IFAE.ES}\,,
Joan SOL\`A
\footnote{Internet addresses: SOLA@IFAE.ES and IFTESOL@CC.UAB.ES}

\vspace{0.25cm}
Grup de F\'{\i}sica Te\`orica\\

and\\

Institut de F\'\i sica d'Altes Energies\\

\vspace{0.25cm}
Universitat Aut\`onoma de Barcelona\\
08193 Bellaterra (Barcelona), Catalonia, Spain\\
\end{center}
\vspace{0.2cm}
\begin{center}
{\bf ABSTRACT}
\end{center}
\begin{quotation}
\noindent
\hyphenation{ana-ly-ses a-na-ly-ses}
We study the correlation of quantum effects on the ratios $R_b$
and $R_c$ within the framework of the MSSM. While in the SM the quantity
$R_b$ is in discrepancy with experiment at the $2\,\sigma$ level
from below, and $R_c$ differs from the experimental result at the
$1.5\,\sigma$ level from above,
the theoretical prediction for both
observables could simultaneously improve in the MSSM,
provided that $\tan\beta$ is large enough
($\tan\beta\sim m_t/m_b$) and there exists
a light supersymmetric pseudoscalar Higgs, and also a light
stop and a light chargino, all of them in the $50\,GeV$ ballpark.
In view of the masses predicted
for these SUSY particles, persistence of the ``$R_b$-$R_c$ crisis'' in the
next run of experiments would not only suggest indirect evidence of SUSY,
but should also encourage direct finding of SUSY at LEP 200. We also point
out the consistency of this picture with other observables and
the intriguing possibility that this $Z$-physics scenario might
allow getting a hint of SUSY at Tevatron through the simple observation of
$t\rightarrow H^{+}b$.
\end{quotation}

\baselineskip=6.5mm  

\newpage


The experimentally measured value of the ratio
\begin{equation}
R_b={\Gamma_b\over\Gamma_h}\equiv {\Gamma (Z\rightarrow b\bar{b})
\over \Gamma (Z\rightarrow {\rm hadrons})}
\label{eq:RbDef}
\end{equation}
has been a source of conflict and of puzzle within the Standard Model\, (SM)
\, in recent times ; and the situation has steadily worsened ever since
the first claimed CDF measurements of the top quark mass, which point towards
a rather high value for this parameter: $m_t=174\pm 16\,GeV$\,\cite{TOPMASS}.
Indeed the present experimental value of the ratio (\ref{eq:RbDef}), which
is accurate to a precision better than  $1\%$, is\,\cite{MARTIN1,MARTIN2}
\begin{equation}
R_b^{\rm exp}=0.2202\pm 0.0020\,,
\label{eq:Rb}
\end{equation}
whereas the theoretical prediction of the SM is found to be (using the
CDF result on $m_t$)\,\cite{BHM}
\begin{equation}
R_b^{SM}=0.2160\pm 0.0006\,.
\label{eq:RbSM}
\end{equation}
With an accuracy better than $3$ parts per mil, the SM prediction is
nonetheless
more than $2$ standard deviations {\it below} the experimental result
 (\ref{eq:Rb}), and it decreases quadratically with the top quark mass due to
a large, negative, vertex contribution to the $b\bar{b}$ mode\,\cite{b1BH1}.
In contrast, $R_b$ is particularly insensitive to the SM Higgs mass and it is
also fairly independent of all sorts of oblique corrections.

In parallel with the conflicting ratio $R_b$, we have another offending
ratio:
\begin{equation}
R_c={\Gamma_c\over\Gamma_h}\equiv {\Gamma (Z\rightarrow c\bar{c})
\over \Gamma (Z\rightarrow {\rm hadrons})}\,.
\label{eq:RcDef}
\end{equation}
Its present experimental value carries a relatively large error ($\sim 6\%$)
\,\cite{MARTIN1,MARTIN2}\,,
\begin{equation}
R_c^{\rm exp}=0.1583\pm 0.0098\,,
\label{eq:Rc}
\end{equation}
but it also defies the prediction of the SM. In fact, in this case the
theoretical result\,\cite{BHM},
\begin{equation}
R_c^{SM}=0.1713\pm 0.0002\,,
\label{eq:RcSM}
\end{equation}
is off by about one and a half standard deviations {\it above}
the experimental value
and it is extremely precise, for it is practically insensitive to the top quark
mass and to the Higgs mass.
 Both ratios (\ref{eq:RbDef}) and (\ref{eq:RcDef})
are independent of $\alpha_s$.

{}From the point of view
of $\Gamma_b$ ($\Gamma_c$), there is an excess (deficit)
of $\sim 8\, MeV$ ($\sim 25\,MeV$) of beauty (charm)
produced in $Z$ decays as compared to SM expectations.
It is thus a challenge to any theory proposing an extension of the SM to
ameliorate the prediction of these observables. In particular,
Supersymmetry (SUSY) and more specifically the Minimal Supersymmetric
Standard Model\,\cite{b1001}, which is supposed to be the most predictive
framework for physics beyond the SM, must be carefully contrasted with
experiment\,\cite{b1002} in all phenomenological fronts
\footnote{Within errors the MSSM is at least as successful as the SM in
the prediction of observables from global fits to all precision data
\,\cite{b1003}, which is certainly {\it not} the case
for the rival composite and technicolour approaches \,\cite{b1203}.}.
Here we shall take the point of
view that there is a ``$R_b-R_c$ crisis'' in the SM and shall explore its
consequences in the MSSM. Failure of the MSSM to improve the theoretical
prediction of the SM, or evidence that it manifestly worsens it,
could be interpreted in the negative sense for SUSY.
However, it should be well borne in
mind that the present status of the experimental information on $R_b$
and $R_c$ is not robust enough to be brandished as a lethal weapon to kill
the MSSM, nor to confirm it. Moreover, consistency of the MSSM
with additional observables will,
of course, be necessary before jumping into conclusions.
 As a matter of fact, the door is still open to
the possibility that the SM itself will be perfectly consistent
with both $R_b^{\rm exp}$ and $R_c^{\rm exp}$
without resorting at all to any form of new physics. We are referring
to the experimental conundrum associated to the process of
b-tagging and its anticorrelation to c-tagging whose resolution might
simultaneously render the $R_b$ and $R_c$ crises non-existent
in the SM\,\cite{MARTIN2,MARTIN3}.
Be as it may, while this technical problem remains unsettled,
we had better prepare the ground to confront the MSSM with the present
and future experimental data on these observables.

Ever since the appearence of the first experimental measurements of $R_b$,
several analyses of supersymmetric radiative corrections to that ratio
have been published in the literature\,\cite{b1GHVR,b1BF5,b1005}. More
recently, $R_b$ has been considered in constrained minimal SUSY
and in specific supergravity models, and in general
to test model building in the framework of supersymmetric
Grand Unified Theories\,\cite{WKK,KimPark,GORDOS}.
Running in parallel with this,
a numerical analysis of complete electroweak radiative corrections to
the full $Z$-width, $\Gamma_Z$, in the general MSSM has been presented
by the authors in Ref.\cite{GJS1}
\,\footnote{That study is based on exact calculations of all
supersymmetric, oblique and non-oblique, one-loop effects on $\Gamma_Z$.
The cumbersome analytical details are presented in Ref.\cite{GJS12}.}; and it
was immediately particularized to the ratio $R_b$ in Ref.\cite{GJS2} whose
notation and definitions we shall adopt hereafter. The aim of the present
letter
is to extent the latter analysis of $R_b$
to situations not explicitly addressed in that  reference and present
the simultaneous prediction of $R_c$ within the MSSM.
For the numerical evaluation we shall borrow
Models I and II as defined in Refs.\cite{GJS1}-\cite{GS}, the first model
being general enough from the phenomenological point of view and
the second one containing the supergravity-based canonical ingredients for
gauge coupling unification. Nonetheless, in the ignorance of the
ultimate unification
theory (if any), and as a means to constrain the profiles of the truly
fundamental
physical theory, we shall not commit ourselves to any particular Yukawa
coupling
unification model, which, if taken seriously enough, should eventually
fit in with the general conditions derived in this study.

In actual practice, we subordinate our combined analysis
of $R_b-R_c$ to the MSSM prediction of  $\Gamma_Z$ \,\cite{GJS1},
which is experimentally bound to lie within the interval\,\cite{MARTIN1}
\begin{equation}
\Gamma_Z^{\rm exp}=2.4974\pm 0.0038\,GeV\,.
\label{eq:GZexp}
\end{equation}
The $\Gamma_Z$-constraint may severely restrict the freedom that we have to
optimize the ratios (\ref{eq:RbDef}) and (\ref{eq:RcDef}).
For any of these ratios we may decompose the MSSM theoretical prediction as
follows:
\begin{equation}
R_{b,c}^{MSSM}=R_{b,c}^{RSM}+ \delta R_{b,c}^{MSSM}\,,
\label{eq:RbMSSM}
\end{equation}
where
\begin{equation}
\delta R_{b,c}^{MSSM}=\delta R_{b,c}^{SUSY}+\delta R_{b,c}^H=
R_{b,c}^{RSM}\,\left({\delta\Gamma_{b,c}^{MSSM}\over \Gamma_{b,c}^{RSM}}
-{\delta\Gamma_h^{MSSM}\over \Gamma_h^{RSM}}\right)
\label{eq:dRbMSSM}
\end{equation}
is the total MSSM departure of these ratios from
the corresponding Reference Standard Model
(RSM) values, $R_{b,c}^{RSM}$, which are identified with (\ref{eq:RbSM}) and
(\ref{eq:RcSM}), respectively \,\cite{GJS2}.
Our calculation of $\delta R_{b,c}^{MSSM}$ includes full
treatment of the genuine
supersymmetric part, $\delta R_{b,c}^{SUSY}$, induced by squarks,
charginos and neutralinos; and also includes full treatment of the
{\it additional}\, Higgs part, $\delta R_{b,c}^H$,
induced by the Higgs sector (charged and neutral) of the MSSM.
In contradistinction to the SM, these extra Higgs-quark
interactions are potentially
significant due to the presence of enhanced Yukawa couplings
involving top and bottom quarks. At the level of the
superpotential they have strengths
\begin{equation}
h_t={g\,m_t\over \sqrt{2}\,M_W\,\sin{\beta}}\;\;\;\;\;,
\;\;\;\;\; h_b={g\,m_b\over \sqrt{2}\,M_W\,\cos{\beta}}\,,
\label{eq:Yukawas}
\end{equation}
shared by the higgsino couplings with the corresponding
quarks and squarks.
In practice, the supersymmetric Yukawa couplings in the mass-eigenstate basis
are interwoven with the gauge couplings in the complete
bottom-stop (sbottom)-chargino (neutralino) interaction
Lagrangian and therefore result in a rather complicated structure
\footnote{ For detailed formulae, see e.g.
eqs.(18)-(19) of Ref.\cite{GHJS}.}.  Numerically we will see that these
couplings are a rather efficient source of non-oblique non-standard
one-loop contributions.
In particular, both the SUSY and the additional Higgs vertex corrections
could be responsible for relatively
important quantum effects on the $Z\rightarrow b\bar{b}$ partial width,
especially if the sparticles are not too heavy.

The signs of all the extra MSSM quantum effects are not coincident.
Thus, on one hand, the SUSY vertex corrections to
$Z\rightarrow b\bar{b}$ are positive whereas those to  $Z\rightarrow c\bar{c}$
are negative.
These signs correspond to the natural regions of parameter space
explored in Refs.\cite{GJS1,GJS2}. For contrived values of the parameters,
they could be different, but we consider it to be unlikely. On the other hand,
the supersymmetric neutral Higgs corrections to
$Z\rightarrow b\bar{b}$ can be either positive or negative
whereas the charged Higgs effects
are always negative. ( Of course, all
Higgs contributions to the $Z\rightarrow c\bar{c}$ partial width
are negligible.) Therefore, in principle we have
more freedom in the MSSM to juggle with the various contributions
in such a way to compensate for the SM ``deficit'' on the $b\bar{b}$
mode while at the same time to cancel out the SM ``surplus'' on the $c\bar{c}$
mode. This extra freedom notwithstanding,
the success of the MSSM should not be viewed as a trivial
adjustment of the parameters; for if the signs described
above would have been just the opposite, then the
supersymmetric contributions would aggravate both the $R_b$ and $R_c$ crises
within the MSSM. Fortunately, the dominant supersymmetric quantum
effects just happen to go in the right direction.

It should be emphasized that we are including the MSSM corrections not only on
the numerators $\Gamma_b$ and $\Gamma_c$ of the ratios (\ref{eq:RbDef}) and
(\ref{eq:RcDef}) but also on all partial widths involved in the denominator
$\Gamma_h$. As a consequence, the SUSY virtual effects on
the $b\bar{b}$ mode (which are positive and constitute
the largest among the SUSY corrections to the total $Z$-width\,\cite{GJS1})
do increase the theoretical value of  both $\Gamma_b$ and $\Gamma_h$.
Thus, as a side effect on $R_c$,
the positive SUSY corrections to $\Gamma_b$
effectively reinforce the negative SUSY contributions to $\Gamma_c$.
All in all, one hopes that
the potential magnitude of the Yukawa couplings (\ref{eq:Yukawas})
allows for a noticeable  shift of $R_b$ up while
at the same time for a shift of $R_c$ a bit down. Thereby
both $R_b^{MSSM}$ and $R_c^{MSSM}$ are expected to be in better agreement
with (\ref{eq:Rb}) and (\ref{eq:Rc}), respectively, than
$R_b^{SM}$ and $R_c^{SM}$.

To assess it quantitatively, we produce in Figs.1-5
the combined plots of $R_b^{MSSM}$ and of
$R_c^{MSSM}$ relevant to our analysis.
We have numerically surveyed the general MSSM parameter space using the
$8$-tuple
procedure devised in Ref.\cite{GJS2}:
\begin{equation}
(\tan\beta, m_{A^0}, M, \mu, m_{\tilde{\nu}}, m_{\tilde{u}},
 m_{\tilde{b}}, M_{LR})\,,
\label{eq:TUPLE}
\end{equation}
from which the whole sparticle spectrum is determined in Model I.
Throughout our numerical analysis, it is understood that
the SUSY parameters in (\ref{eq:TUPLE})
will be scanned in the wide intervals given in eq.(14) of Ref.\cite{GJS2} under
certain conditions to be specified in each case.
Notice that because
of the $8$-dimensional nature of the parameter space, all our numerical
searches and optimizations are highly CPU-time demanding. Just to get an idea,
the working out of our figures took a few
hundred hours of net CPU-time in an IBM (RS 6000, 390/3BT) and in an
``$\alpha$'' (DEC 3000, 300/AXP).

Our search for admissible points is optimized by fixing the pseudoscalar
mass, $m_{A^0}$, to a few light values bordering the phenomenological lower
bounds\,\cite{b1PR1,b1STE}.
In a regime of high $\tan\beta$ these
choices insure that the positive contribution to $R_b^{MSSM}$
from the neutral Higgs sector of the MSSM is large enough to
override the negative contribution from the charged
Higgses (Cf. Fig.3 of Ref.\cite{GJS2}). For heavy
pseudoscalar masses this situation is
not possible; and, as we have shown in the latter reference,
the position of $R_b$ becomes far less comfortable in the MSSM.
Further optimization of $R_b^{MSSM}$ (i.e. additional positive contributions
to that quantity)
is achieved by searching over regions of the SUSY parameter space
where at least one chargino and one stop have a mass as close as possible
to their present lower phenomenological
bounds\,\cite{b1PR1,b1STE,b1SJA}.
Thus, to start with, we restrict the search of points (\ref{eq:TUPLE}) within
the
subspace
\begin{equation}
45\,GeV<m_{\tilde{t}_1}<60\,GeV\,,\ \
48\,GeV<M_{\Psi^{\pm}_1}<60\,GeV\,, \ \ M_{\Psi_1^0}>20\,GeV\,,
\label{eq:OPTIM1}
\end{equation}
where $m_{\tilde{t}_1}$, $M_{\Psi^{+}_1}$, $M_{\Psi^{0}_1}$ are the
masses of the lightest stop, chargino and neutralino, respectively.
(The other stop-chargino-neutralino mass eigenvalues can, of course, be
heavier.)
This will insure at the same time a sizeable negative contribution to
$R_c^{MSSM}$.

{}From Fig.1a we learn that to restore the ratio $R_b^{MSSM}$ within
$1\,\sigma$ of
$R_b^{\rm exp}$ we have to require
$\tan\beta\geq 22$.
Furthermore, since $R_b^{MSSM}$ is more yielding than $R_c^{MSSM}$ (Fig.1b),
we have focused our work on optimizing
$R_c^{MSSM}$\,; thus our figures actually show the simultaneous
solution curves for $R_b^{MSSM}$ corresponding to the best solution curves for
$R_c^{MSSM}$ obtained in the aforementioned intervals of parameter space.

Our optimum curves concentrate around the lightest values of the
stop and chargino masses in the range (\ref{eq:OPTIM1}). Although this
could be expected, we did not try to fix the masses  $m_{\tilde{t}_1}$ and
$M_{\Psi^{+}_1}$ beforehand, but rather
we let our code to search for the optimum values automatically within the
parameter
subspace under consideration. (Decoupling of SUSY in the asymptotic regime
should
not preclude the possibility of interesting local behaviours.)

In contrast to $R_b^{MSSM}$,
it turns out that it is not possible to drag $R_c^{MSSM}$ into the
experimental range at the strict $1\,\sigma$ level.
However, simple inspection of Figs.1a-1b shows that
for any value of $\tan\beta$ that makes $R_b^{MSSM}$ compatible with
$R_b^{\rm exp}$ at $1\,\sigma$, makes $R_c^{MSSM}$ compatible with
$R_c^{\rm exp}$ at $1.25\,\sigma$. Therefore, on condition that
\footnote{This is still well below the approximate
perturbative limit $\tan\beta\sim 70$.
 Incidentally we note that the large $\tan\beta$ region is
favoured by recent $t-b-\tau$ Yukawa coupling $SO(10)$ unification
models\,\cite{SO10}.}
\begin{equation}
\tan\beta\stackrel{\scriptstyle >}{{ }_{\sim}}20\,,
\label{eq:tan1}
\end{equation}
it is possible to simultaneously solve the ``$R_b$ crisis'' at
 $1\,\sigma$ and the ``$R_c$ crisis'' at $1.25\,\sigma$ within the MSSM.
To appreciate the sensitivity of the curves in Fig.1 to the variation of
the parameters,  we fix e.g. $m_{A^0}=40\,GeV$ and sample our best solution
curve
for $R_c^{MSSM}$ over the range (\ref{eq:OPTIM1}).
The result is represented in the form of a narrow band in Fig.2b, whose
darkened part is compatible with $R_c^{\rm exp}$ to within $1.25\,\sigma$.
The one-to-one map of this darkened
region onto the $(R_b^{MSSM},\tan\beta)$-plane is the other
darkened band shown in Fig.2a, part of which is excluded by the upper
and lower $1\,\sigma$ cuts on $R_b^{\rm exp}$.

Let us remark that the  $\Gamma_Z$-constraint mentioned above is innocuous
in Figs.1-2, due to the small SUSY correction to the $Z$-width
within the parameter subspace (\ref{eq:OPTIM1}). The smallness of the
correction
stems from the large, negative, vacuum polarization
effects on $\Gamma_Z$ from chargino-neutralinos
in that region of parameter space (Cf. Fig.1 of Ref.\cite{GJS1}) which
nevertheless cancel out to a large extent in $R_{b,c}$.

In Figs.3a-3b we display the best (candidate) solution curves for
$R_c^{MSSM}$ in correspondence with the simultaneous solution curves
for $R_b^{MSSM}$ when the chargino-stop spectrum is scanned in
the intermediate mass region up to the LEP 200 discovery range:
\begin{equation}
60\,GeV<m_{\tilde{t}_1}, M_{\Psi^{\pm}_1}<90\,GeV\,.
\label{eq:OPTIM2}
\end{equation}
Again, our code projects the best solution curves for the
lightest values of the
stop and chargino masses in this range. However, since
this time all the supersymmetric vacuum polarization
corrections are positive, and therefore add up to the leading (positive)
vertex contributions, the effect of the $\Gamma_Z$-constraint becomes patent:
It cuts-off the (candidate) solution curves and prevents them from
exiting the experimentally allowed region. In the same conditions as before,
a simultaneous solution of the ``$R_b-R_c$ crisis'', however, does exist for
\begin{equation}
\tan\beta\stackrel{\scriptstyle >}{{ }_{\sim}}25\,.
\label{eq:tan2}
\end{equation}
We have checked that in running through the values of the interval
(\ref{eq:OPTIM2})
from lower to higher masses $m_{\tilde{t}_1}$ and $ M_{\Psi^{\pm}_1}$,
the cut-off effect from the $\Gamma_Z$-constraint
trims away a larger and larger
portion of the optimum solution curves.

The critical cut-off situation shown in Figs.4a-4b corresponds
to a numerical search
in the vicinity of the LEP 200 unaccessible range
\begin{equation}
90\,GeV<m_{\tilde{t}_1}, M_{\Psi^{\pm}_1}<110\,GeV\,.
\label{eq:OPTIM3}
\end{equation}
Here both $R_b^{MSSM}$ and $R_c^{MSSM}$ find themselves in deep water.
Indeed, these curves are severely cut-off; and
whereas $R_b^{MSSM}$ scarcely penetrates into
the experimental domain of $R_b^{\rm exp}$ at $1\,\sigma$, $R_c^{MSSM}$
is unable to reach $R_c^{\rm exp}$ at all, not even at $1.25\,\sigma$.
Hence, in the mass interval under consideration
a simultaneous solution to the ``$R_b-R_c$ crisis'' within the MSSM
does not exist for any value
of $\tan\beta$, unless the error on $R_c^{\rm exp}$ is extended up to
$1.5\,\sigma$,
i.e. up to the compatibility range of the SM itself.
Notice that even in this case the
MSSM is in better shape than the SM, for the MSSM could still be
marginally consistent with $R_b$ at $1\,\sigma$
(for $\tan\beta\stackrel{\scriptstyle >}{{ }_{\sim}}30$)
whereas the SM would be at variance with experiment
by more than $2\,\sigma$.

The transition from the ``free regime'' of Figs.1-2
into the ``cut-off regime'' of Figs.3-5 turns on for
$m_{\tilde{t}_1}, M_{\Psi^{\pm}_1}>55\,GeV$,
reaching a maximum somewhere beyond the LEP 200
unaccessible range (\ref{eq:OPTIM3})
and then becoming again less and less severe as long as
the sparticle masses become
effectively decoupled. As stated, this to-and-fro behaviour
is due to a balance
between the oblique and non-oblique corrections and to the
fact that the formers (latters) are leading effects on
$\Gamma_Z$ ($R_{b,c}^{MSSM}$)
but virtually cancel in $R_{b,c}^{MSSM}$.

Finally, we display in Figs.5a-5b the case corresponding to
very heavy sparticles, where the ratios $R_{b,c}^{MSSM}$ recover from the
critical cut-off behaviour.
This set-up corresponds in good approximation to what we have termed Model II.
Here the effect of the $\Gamma_Z$-constraint is in fact not too harmful, for
the
supersymmetric contributions (oblique and non-oblique)
are very small and care is needed only to control the Higgs effects.
The situation depicted in Fig.5 actually corresponds to the
asymptotic cut-off regime where the various sparticles are infinitely heavy
\footnote{Slight differences with respect to Fig.5 of Ref.\cite{GJS2} are due
to
updating of $R_b^{\rm exp}$ in the present study.}.
We see that in this asymptotic regime
a simultaneous MSSM solution exists for $R_b$ at $1\,\sigma$ and
for $R_c$ at $1.25\,\sigma$ provided
\begin{equation}
\tan\beta\stackrel{\scriptstyle >}{{ }_{\sim}}45\,,
\label{eq:tan3}
\end{equation}
i.e. for significantly larger values of $\tan\beta$ than in the light and
intermediate SUSY cases, eqs.(\ref{eq:tan1}), (\ref{eq:tan2}).
Needless to say, for sparticle masses well beyond the LEP 200 discovery limit
and at the same time a pseudoscalar mass $m_A>70\,GeV$,
the position of $R_{b,c}$ in the MSSM would be as untenable as in the SM.
It is nevertheless quite remarkable that
the sole presence of a light pseudoscalar may greatly
alleviate the ``$R_b-R_c$ crisis''
at the modest expense of a large value of $\tan\beta$. This feature, which is
automatic in the MSSM Higgs sector, could also be achieved in general
two-Higgs-doblet-models, but only after a suitable tuning of the parameters.

In Ref.\cite{GJS2} we showed
that $R_b^{\rm exp}$ could tolerate (at $1\,\sigma$)
a SUSY spectrum in the vicinity of the LEP 200 unaccessible range
(\ref{eq:OPTIM3}), so long as one keeps
a light pseudoscalar Higgs
and a large value of $\tan\beta$. From the present study we
realize that this is no longer possible if we want at the same time to match up
$R_c^{MSSM}$ with $R_c^{\rm exp}$ to an accuracy better than
$1.25\,\sigma$.  Admittedly,
the MSSM achievement on this ratio
may not be too spectacular; after all the experimental
error on $R_c$ is much larger than that on $R_b$ and therefore the experimental
situation of $R_c$ is still loose enough to undergo potentially important
changes
in the near future.
All the same, the improvement of the prediction of $R_b$ in the MSSM
is clear-cut and we can build on the fact that the corresponding
effect on $R_c$ has at least the right trend and it can be
quantitatively acceptable in situations like the ones depicted
in Figs.1-3 and 5.

The nature of the solutions in these figures is however rather different.
In fact, although the comfortable
solution in Figs.1-2 is free from the $\Gamma_Z$-constraint
and prefers the lightest possible values for some sparticle masses, as a
drawback
it is confined to the narrow interval (\ref{eq:OPTIM1}). In contrast, the
solution
in Figs.3 and 5, in spite of being a cut-off solution, it is perfectly
compatible with a relatively light or a very heavy (decoupled)
sparticle spectrum.
Thus, in the very end, what is needed from the MSSM is either a light
or a heavy SUSY spectrum (not an intermediate one!),
together with a light pseudoscalar and a large value of $\tan\beta$.
With all these ingredients, the MSSM is able to cook a fairly satisfactory
resolution of the ``$R_b-R_c$ crisis'' which encourages
LEP 200 to find at least a CP-odd (and a CP-even) supersymmetric Higgs,
and in favourable circumstances (Figs.1-2) even a stop and a chargino.

It is worthwhile to note, in passing, a couple of interesting
consistency facts of our results with the status of other observables:
i) One fact is that the above picture
fits pretty well with the theoretical requisites for the branching ratio
of $b\rightarrow s\,\gamma$
to be compatible with experiment within the MSSM. Indeed, a light CP-odd Higgs
at large $\tan\beta$ is perfectly
allowed by $B (b\rightarrow s\,\gamma)$\,\cite{BSGAMMA1};
 besides, both a light chargino and a light
stop at large $\tan\beta$ are precisely needed to coexist peacefully with
a not too heavy charged Higgs\,\cite{BSGAMMA2}, i.e. such that to produce a
net global radiative correction preserving the CLEO bounds\,\cite{CLEO};
ii) The other consistency fact is that this scenario
could also help to cure the bold $3\,\sigma$ discrepancy between
the low energy and high energy determinations of $\alpha_s(M_Z)$ from global
fits to all indirect precision data within the SM.
As remarked  in Ref.\cite{LANGACK}, any increase of $\Gamma_h$
coming from physics beyond the
SM would be wellcome in this respect, for it would diminish the high
energy (lineshape) value of $\alpha_s(M_Z)$ obtained
from $R=\Gamma_h/\Gamma_l$.
Now, in our particular MSSM scenario, the balance resulting from
the full set of extra electroweak quantum effects on $\Gamma_h$
ends up with a net positive contribution to that quantity; and, what is more,
if we place ourselves in the very same parameter region that
we have been exploiting to ameliorate the theoretical predictions of $R_b$ and
$R_c$, we automatically obtain
the necessary $4$ per mil enhancement of $\Gamma_h$ to solve the
``$\alpha_s(M_Z)$ crisis'' too\footnote{For a detailed study of this
issue, see Ref.\cite{ALFACRI}}.
Cards are, therefore,
laid down on the table just awaiting for the next round of experiments.

A final comment is in order concerning the demand
of the present analysis, and of previous analyses
\,\cite{b1GHVR}-\cite{GJS2}, for light SUSY particles,
and more specifically for a light stop, a light chargino--and a fortiori a
light neutralino.
It obviously suggests that we should see some
supersymmetric top quark decay at the next Tevatron run. Thus, even
barring the (still open) possibility
of light gluinos, we should expect electroweak decays like
$t\rightarrow\tilde{t}_1+ \Psi^{0}_1$ and
$t\rightarrow\tilde{b}_1+\Psi^{+}_1$ if there is a not too heavy
sbottom, $\tilde{b}_1$.
Interesting as they can be, however,
all these modes involve genuine SUSY particles and so
their actual detection may require some additional effort to tag
unconventional final states. Alternatively-- and
as a distinctive feature of the present analysis--it is amusing to notice
that the conditions to solve the ``$R_b-R_c$ crisis'' within
the MSSM suggest that another, less exotic, decay mode of the
top quark should be $t\rightarrow H^{+}b$.
In fact, we have seen that the correlation $R_b-R_c$ demands
not only a light chargino and a light stop but also a light value
for the pseudoscalar Higgs mass around $50\,GeV$. Thus,
from the well-known MSSM Higgs mass relations\,\cite{HUNTER}, it follows that
$m_{H^{+}}\sim 100\,GeV$ and therefore $t\rightarrow H^{+}b$
is expected to be not too much suppressed by phase space
with respect to the standard decay mode $t\rightarrow W^{+}b$\,
\footnote{Potentially important electroweak and strong
supersymmetric virtual effects on this decay
have recently been recognized in the literature\,\cite{GHJS,LYH,HJJS}.}.
Furthermore, the marked preference that $R_b-R_c$ has
for the regime of large values of $\tan\beta$ suggests that the two
decay widths \,\cite{BERNR}
\begin{equation}
\Gamma(t\rightarrow W^{+}b)=
{G_F m_t^3\over 8\pi\sqrt{2}}\,
\left(1-{M_W^2\over m_t^2}\right)^2\,\left(1+2{M_W^2\over m_t^2}\right)
\label{eq:treeW}
\end{equation}
and
\begin{equation}
\Gamma(t\rightarrow H^{+}b)={G_F m_t^3\over 8\pi\sqrt{2}}\,
\left(1-{m_{H^{+}}^2\over m_t^2}\right)^2\,
\left[{m_b^2\over m_t^2}\,\tan^2\beta+\cot^2\beta\right]
\label{eq:treeH}
\end{equation}
can be comparable.
Last but not least,
from the fact that $\tan\beta$ could be so large, we expect an additional
bonus:
namely, that a supersymmetric
charged Higgs should most likely decay into $\tau$-lepton and neutrino,
rather than into charm-strange quark jets.
Indeed this immediately follows from
\begin{equation}
\Gamma(H^{+}\rightarrow\tau^{+}\nu_{\tau})=
{G_F m_{\tau^{+}}^2\,m_{H^{+}}\over 4\pi\sqrt{2}}\,\tan^2\beta
\label{eq:treetau}
\end{equation}
and
\begin{equation}
\Gamma(H^{+}\rightarrow c\bar{s})=
{3 G_F m_c^2\,m_{H^{+}}\over 4\pi\sqrt{2}}\,
\left[{m^2_s\over m^2_c}\tan^2\beta+\cot^2\beta\right]\,.
\label{eq:treecs}
\end{equation}
 The fact that the few Tevatron events collected on the top quark do not
make such a distinction, probably means that they correspond to the standard
decay mode. Nevertheless, a better statistics
and an appropriate experimental search in the future might
come across some $\tau$-lepton final
states whose parent particle is not just a $80\,GeV$ good-natured $W$-boson but
a ${\cal O}(100)\,GeV$ fully-fledged charged Higgs
\footnote{Studies from the LHC collaborations\,\cite{ATLAS}
show that for $m_{H^{\pm}}<130\,GeV$ and large $\tan\beta$
it is possible to identify
the $H^{+}\rightarrow\tau^{+}\nu_{\tau}$ decay on account of the observed
excess of
events with one isolated $\tau$ as compared to events with an additional
lepton\,\cite{MBOSMAN}.}.
 In principle there is no
a priori reason for $\Gamma(t\rightarrow H^{+}b)$ to be competitive with
$\Gamma(t\rightarrow W^{+}b)$, nor for a non-supersymmetric charged Higgs to
decay
most likely into the $\tau$-lepton mode rather than into the hadronic mode.
Nonetheless for a supersymmetric charged Higgs we do have, in the light of the
MSSM quantum effects
on $R_b-R_c$, a reasonable motivation to believe in such a scenario. So,
at the end of the day, the message for the experimentalists
could be something like:
find (or manage to find!)  $H^{+}\rightarrow \tau^{+}\,\nu_{\tau}$
at Tevatron or at LHC ( however difficult as $\tau$-tagging can be in this
context)
and you might be discovering SUSY!.

\newpage


{\bf Acknowledgements}:
 One of us (JS) thanks M. Mart\'\i nez for reading the
manuscript and for helpful conversations on the experimental status of
$R_b$ and $R_c$. He is also thankful to M. Bosman for an enlightening
discussion on the Higgs detection possibilities at Tevatron and LHC.
Interest on our work by W. de Boer and D. Finnell is also gratefully
acknowledged. This work has been partially supported by CICYT under project
No. AEN93-0474. The work of DG has also been financed by a grant of the
Comissionat per a Universitats i Recerca, Generalitat de Catalunya.

\vspace{1.5cm}



\vspace{2cm}
\begin{center}
\begin{Large}
{\bf Figure Captions}
\end{Large}
\end{center}
\begin{itemize}
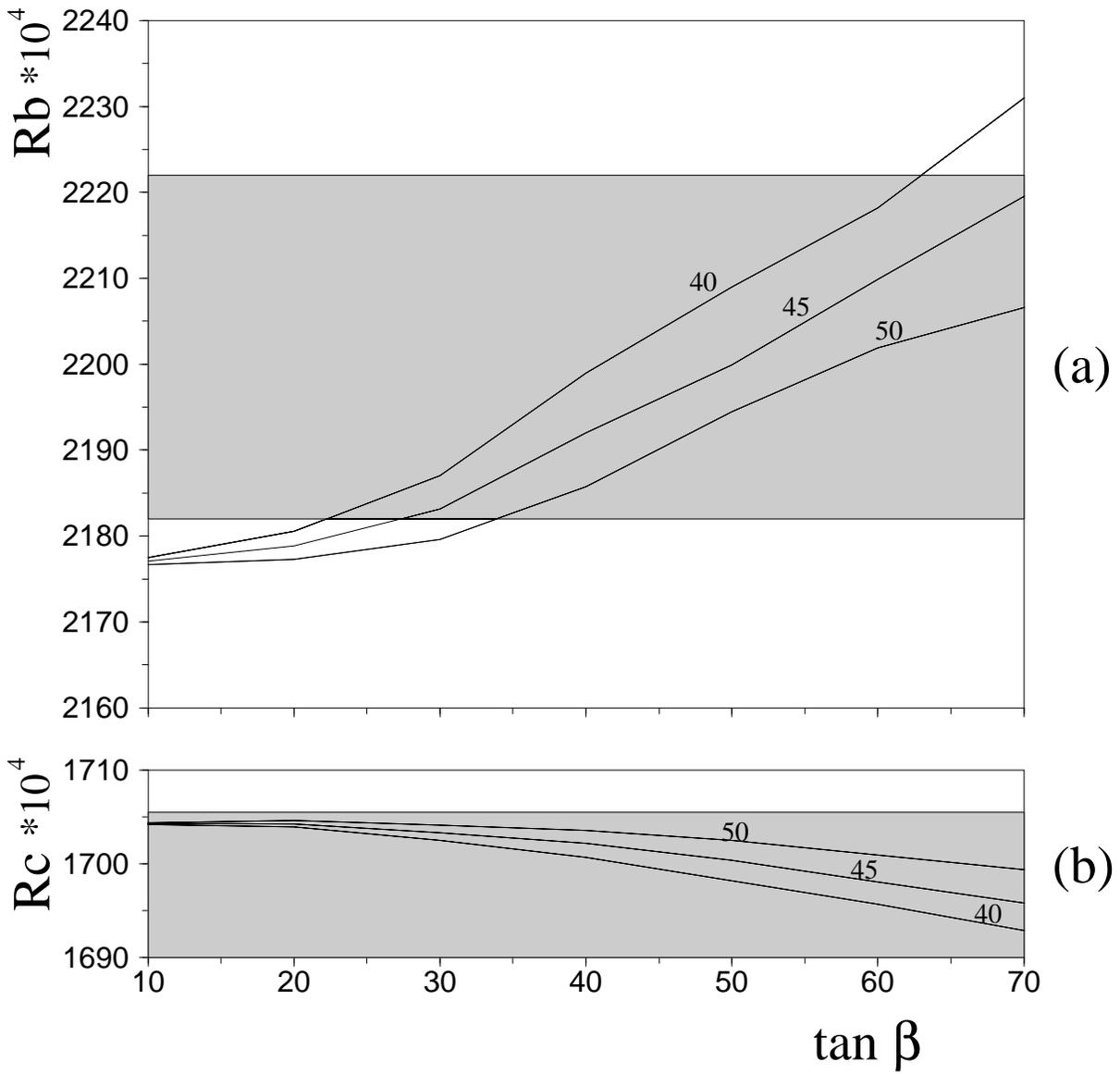
\item{\bf Fig.1} (a) $R_b^{MSSM}$ as a function of $\tan\beta$ for three light
pseudoscalar masses $m_{A^0}=40,45,50\,GeV$ (curves from top to bottom),
in correspondence with (b) the best solution curves
for $R_c^{MSSM}$ (from bottom to top).
The SUSY spectrum, eq.(\ref{eq:TUPLE}), was
scanned in the light chargino-stop mass range (\ref{eq:OPTIM1}).
The shaded area
in (a) corresponds to $R_b^{\rm exp}$ within $1\sigma$ whereas that in (b)
corresponds to the upper part of  $R_c^{\rm exp}$ within $1.25\sigma$.
We have taken $m_b=5\,GeV$ and $m_t=174\,GeV$.

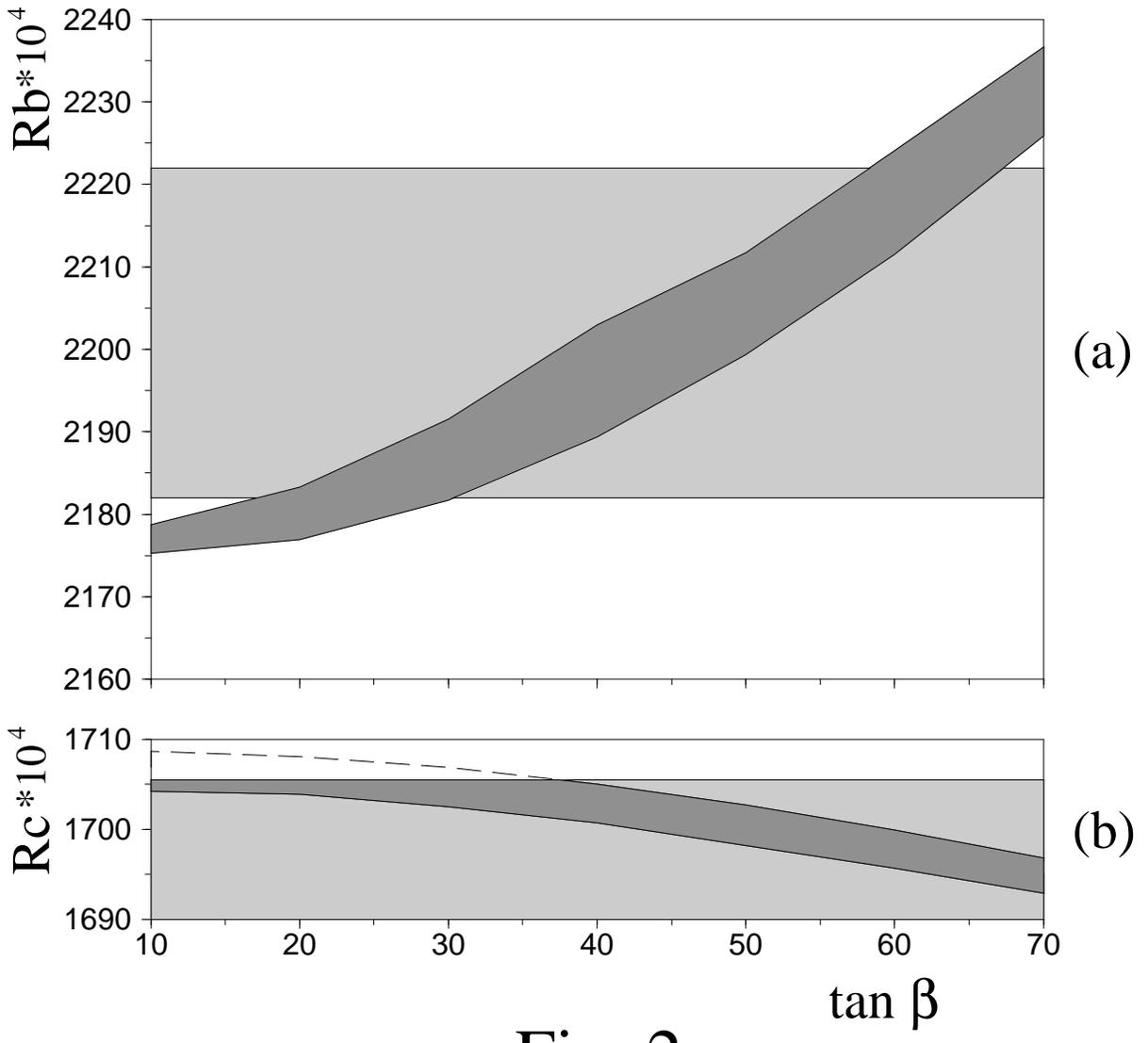
\item{\bf Fig.2} Sensitivity of the solution curve $m_{A^0}=40\,GeV$ of Fig.1
to a sweep of the parameters across the intervals (\ref{eq:OPTIM1}).
The narrow darkened bands
in (a) and in (b) correspond to $R_b^{MSSM}$ and $R_c^{MSSM}$, respectively,
and are in one-to-one correspondence as explained in the text.

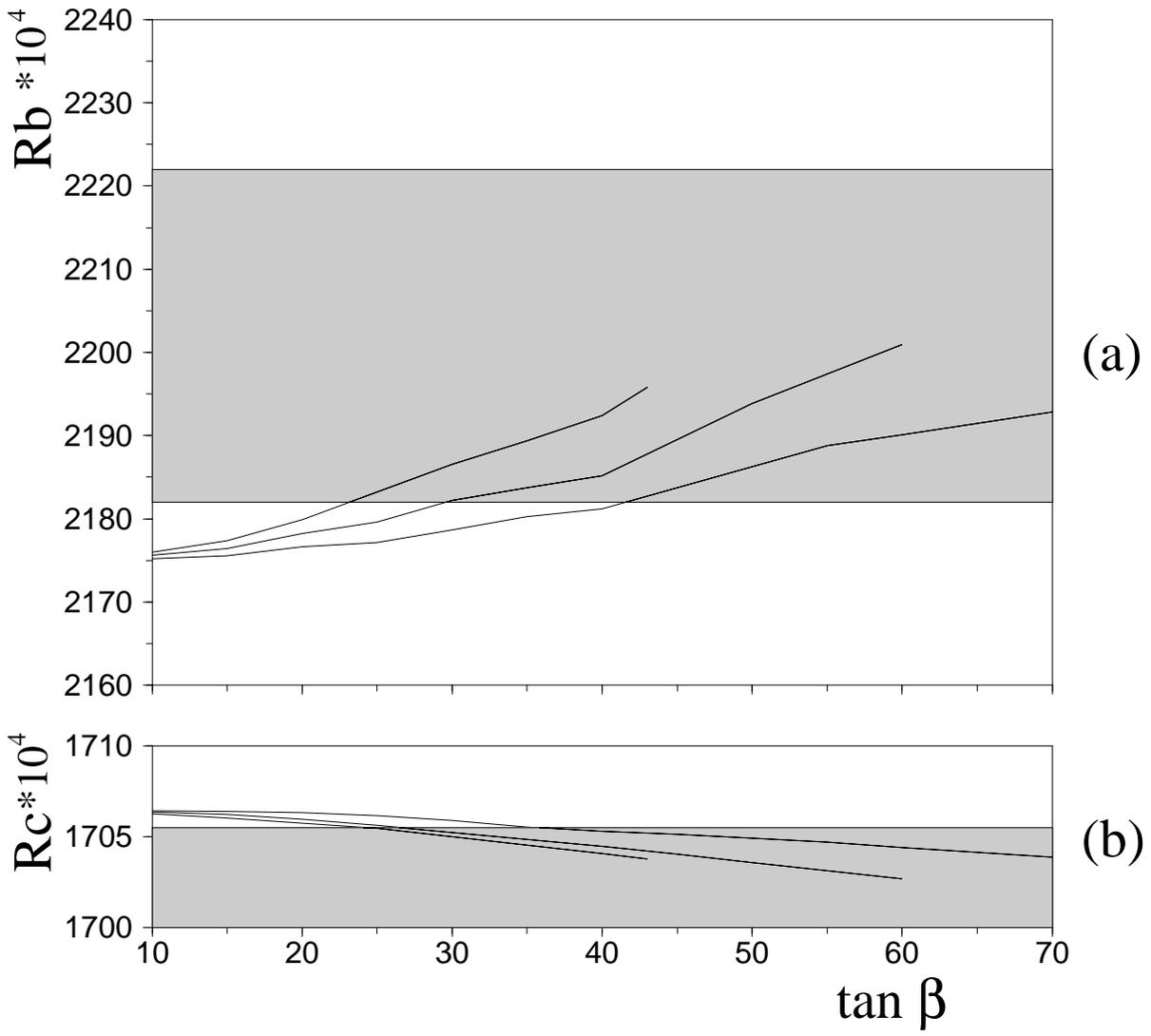
\item{\bf Fig.3} As in Fig.1, but for a parameter survey in the
intermediate chargino-stop region (\ref{eq:OPTIM2}), which reaches up to the
LEP 200 discovery range.

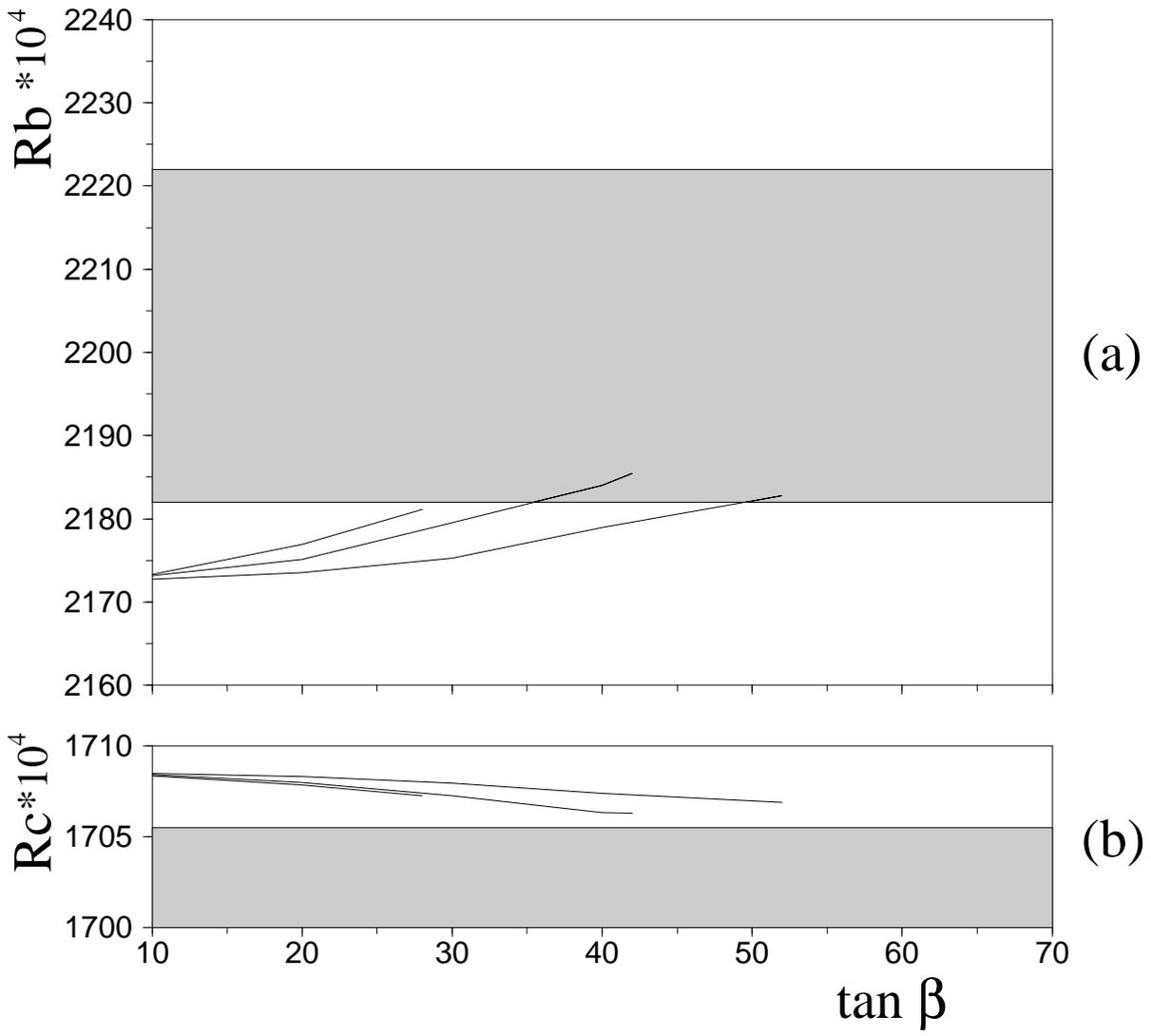
\item{\bf Fig.4} As in Fig.1, but for a parameter survey in the vicinity of the
LEP 200 unacessible region (\ref{eq:OPTIM3}).

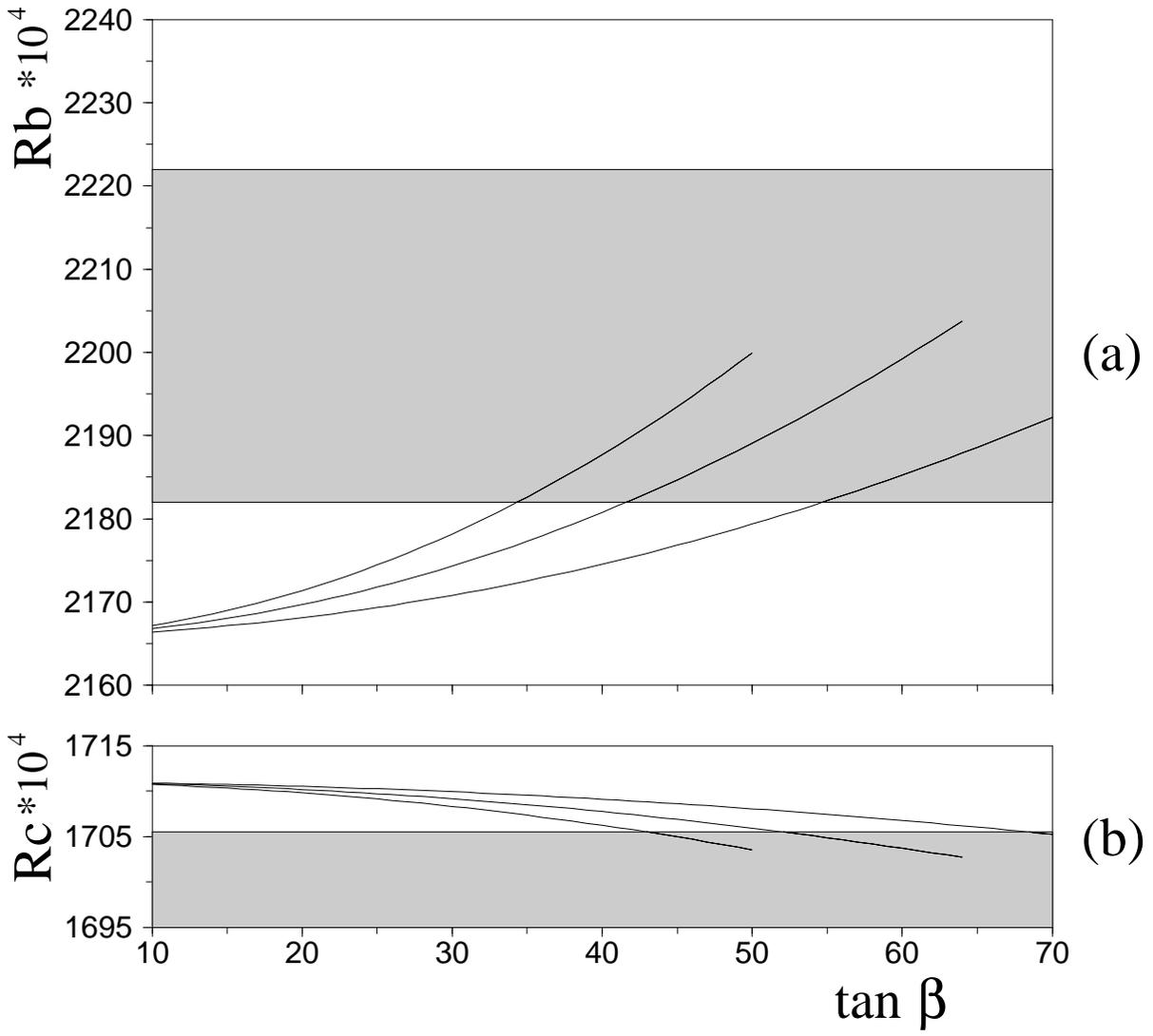
\item{\bf Fig.5}  As in Fig.1, but for a sparticle spectrum fully decoupled.

\end{itemize}


\begin{thebibliography}{99999999}

\bibitem{TOPMASS}
F. Abe et al. (CDF Collab.), Phys. Rev. Lett. {\bf 73} (1994) 225; {\it ibid}
Phys. Rev. {\bf D50} (1994) 2966.
\bibitem{MARTIN1}
The LEP Collaborations ALEPH, DELPHI, L3, OPAL and the LEP Electroweak Working
Group, CERN-PPE/93-157 and Contribution to the 27th International Conference
on High Energy Physics, Glasgow, Scotland, July 1994 (to appear in the
Proceedings).
\bibitem{MARTIN2}
 M. Mart\'\i nez,\, {\it Results at LEP on electroweak parameters},
preprint IFAE-UAB/94-02, September 1994 ( to appear in: Proc. of the
XXII International Meeting on Fundamental Physics: ``The standard model
and beyond'', Jaca, Spain, February 1994); and {\it Precision tests of the
standard model}, preprint IFAE-UAB/95-01, January 1995.
\bibitem{BHM}
We have used the upgraded version of the computer code BHM,
by G. Burgers, W. Hollik and M. Mart\'{\i}nez; M. Consoli,
W. Hollik and F. Jegerlehner: Proc. of the
{\it Workshop on $Z$ Physics at LEP1},
CERN 89-90, Sept. 1989, ed. G. Altarelli et al., Vol.1, p.7; G. Burgers,
F. Jegerlehner, B. Kniehl and J. K\"uhn: the same Proc. Vol.1, p.55.
\bibitem{b1BH1}
J. Bernabeu, A. Pich and A. Santamaria, Phys. Lett. {\bf B200} (1988) 569;
W. Beenaker and W. Hollik, Z. Phys. {\bf C40} (1988) 141; A. Akhundov, D.
Bardin
and T. Riemann, Nucl.Phys. {B276} (1986) 1.
\bibitem{b1001}
H. Nilles, Phys. Rep. {\bf 110} (1984) 1; H. Haber and G. Kane, Phys. Rep.
{\bf 117} (1985) 75;
 A. Lahanas and D. Nanopoulos, Phys. Rep. {\bf 145} (1987) 1;
 See also the exhaustive reprint collection {\it Supersymmetry}
(2 vols.), ed. S. Ferrara (North Holland/World Scientific, Singapore, 1987);
W. de Boer,\, {\it Grand unified theories and supersymmetry in particle physics
and cosmology}, U. Karlsruhe preprint IEKP-KA-94-01, February 1994\, (to be
published in ``Progress in Particle and Nuclear Physics'').
\bibitem{b1002}
Proc. of the workshop:
{\it Ten Years of SUSY Confronting Experiment}, ed. J. Ellis, D.V. Nanopoulos
and A. Savoy-Navarro, CERN, September 1992, CERN-TH.6707/92-PPE/92-180.
\bibitem{b1003}
J. Ellis, G.L. Fogli and E. Lisi, Phys. Lett. {\bf B333} (1994) 118;\,
{\it ibid} Nucl.Phys. {\bf B393} (1993) 3;
J. Erler and P. Langacker, in\,\cite{LANGACK}.
\bibitem{b1203}
J. Ellis, G.L. Fogli and E. Lisi, Phys. Lett. {\bf B343} (1995) 282;
J. Erler and P. Langacker, in\,\cite{LANGACK}.
\bibitem{MARTIN3}
M. Mart\'\i nez, private conversation.
\bibitem{b1GHVR}
A. Djouadi, G. Girardi, C. Verzegnassi, W. Hollik and F.M. Renard, Nucl. Phys.
{\bf B349} (1991) 48.
\bibitem{b1BF5}
M. Boulware and D. Finnell, Phys. Rev. {\bf D44} (1991) 2054.
\bibitem{b1005}
 G. Altarelli, R. Barbieri and F. Caravaglios,
Phys. Lett. {\bf B314} (1993) 357; {\it ibid} Nucl. Phys.{\bf B405} (1993) 3;
 G. Altarelli, talk at the
{\it Tennessee International Symposium on
Radiative Corrections}, Gatlinburg, Tennessee, June 27-July 1, 1994 (to appear
in the Proceedings).
\bibitem{WKK}
J.D. Wells, C. Kolda and G.L. Kane, Phys. Lett. {\bf B338} (1994) 219;
G.L. Kane, C. Kolda, L. Roszkowski and J. D. Wells, Phys. Rev. {\bf D49}
(1994) 6173 and {\bf D50} (1994) 3498.
\bibitem{KimPark}
J.E. Kim and G.T. Park,\,  $\epsilon_b$\,{\it constraints on the minimal
$SU(5)$
and $SU(5)\times U(1)$ supergravity models},
preprint SNUTP 94-66, August 1994; X. Wang, J.L. Lopez and
D.V. Nanopoulos,\, {\it $R_b$ in supergravity models},
preprint CERN-TH.7553/95, January 1995.
\bibitem{GORDOS}
M. Carena and C. Wagner,\, {\it Higgs and supersymmetric particle signals at
the infrared fixed point of the top quark mass},
preprint CERN-TH.7393/94, August 1994.
\bibitem{GJS1}
D. Garcia, R.A. Jim\'enez and J. Sol\`a,\, {\it Supersymmetric electroweak
renormalization of the $Z$-width in the MSSM}, preprint UAB-FT-343,
September 1994 (hep-ph/9410310) (Phys. Lett. B, in press).
\bibitem{GJS12}
D. Garcia, R.A. Jim\'enez and J. Sol\`a,\, {\it The width of the $Z$ boson in
the
MSSM}, preprint UAB-FT in preparation.
\bibitem{GJS2}
D. Garcia, R.A. Jim\'enez and J. Sol\`a,\, {\it Full electroweak quantum
effects on
$R_b$ in the MSSM}, preprint UAB-FT-344, September 1994
(hep-ph/9410311)  (Phys. Lett. B, in press).
\bibitem{GS}
D. Garcia and J. Sol\`a, Mod. Phys. Lett. {\bf A9} (1994) 211.
\bibitem{GHJS}
D. Garcia, W. Hollik, R.A. Jim\'enez and J. Sol\`a,
Nucl. Phys. {\bf B427} (1994) 53.
\bibitem{b1PR1}
Aleph Collab., Phys. Rep. {\bf 216} (1992) 254; Phys. Rev.{\bf D50} (1994)
1369.
\bibitem{b1STE}
P. Abreu, et al. (DELPHI Collab.), Phys. Lett. {\bf B247} (1990) 148;
O. Adriani, et. al. (L3 Collab.), Phys. Rep. {\bf 236} (1993) 1.
\bibitem{b1SJA}
T. Kon and T. Nonaka, Phys. Lett. {\bf B319} (1993) 355; H. Baer, J. Sender
and X. Tata, Phys. Rev {\bf D50} (1994) 4517.
\bibitem{SO10}
R. Rattazzi, U. Sarid and L.J.Hall, preprint SU-ITP-94-15, May 1994, to appear
in: Proc. of the {\it 2nd Workshop on Yukawa Couplings and the Origin of
Mass}, Gainesville, Florida, February 1994.
\bibitem{LYH}
J.M. Yang and C.S. Li, Phys. Lett.{\bf B320} (1994) 117.
\bibitem{HJJS}
W. Hollik, R.A. Jim\'enez, C. J\"unger and J. Sol\`a,\, {\it Strong
supersymmetric
quantum effects on the top quark width}, Karlsruhe preprint KA-TP-1-1995
and Universitat Aut\`onoma de Barcelona preprint UAB-FT-357, January 1995.
\bibitem{BSGAMMA1}
M.A. Diaz, Phys. Lett. {\bf B304} (1993) 278.
\bibitem{BSGAMMA2}
R. Garisto and J.N. Ng, Phys. Lett. {\bf B315} (1993) 372
(See references therein for early works on this issue).
\bibitem{CLEO}
E. Thorndike et al. (CLEO Collab.), Phys. Rev. Lett. {\bf 71} (1993) 674.
\bibitem{LANGACK}
J. Erler and P. Langacker,\, {\it Implications of high precision experiments
and the CDF top quark candidates},
preprint UPR-0632T, October 1994; M. Shifman,\, {\it Determining $\alpha_s$
from measurements at Z: how nature prompts us about new physics}, preprint
TPI-MINN-94/42-T, December 1994.
\bibitem{ALFACRI}
D. Garcia and J. Sol\`a,\, {\it Matching the low and high energy
determinations of $\alpha_s(M_Z)$ in the MSSM}, preprint UAB-FT, in
preparation.
\bibitem{HUNTER}
J.F. Gunion, H.E. Haber, G. Kane and S. Dawson, {\it The Higgs hunter's guide}
(Addison-Wesley, New York, 1990).
\bibitem{BERNR}
W. Bernreuther et al., in
Proc. of the Workshop on $e^+e^-$ Collisions at 500 GeV: The Physics
Potential, Hamburg, 1991, ed. P.M. Zerwas; preprint DESY 92-123A, 1992.
\bibitem{ATLAS}
Atlas Collab., {\it Atlas technical proposal for a general-purpose pp
experiment
at the Large Hadron Collider at CERN}, preprint CERN/LHCC/94-43, December 1944,
pp. 245-248.
\bibitem{MBOSMAN}
M. Bosman, private conversation.

\end{thebibliography}
\end{document}